\documentclass{CVIS}
\usepackage{amsmath,bm}
\usepackage{graphicx}% Include figure files
\usepackage{dcolumn}% Align table columns on decimal point
\usepackage{subcaption,soul}
\usepackage{caption}
\usepackage[numbers,sort&compress,sectionbib]{natbib}
\usepackage[pagebackref=flase,breaklinks=true,letterpaper=true,colorlinks=true,linkcolor = blue, urlcolor = black,citecolor=black,bookmarks=true]{hyperref}
\usepackage{url}
\usepackage{mleftright}
\usepackage{multirow}
\usepackage{booktabs}
\usepackage{titlesec}

\title{Semi-supervised Anomaly Detection using AutoEncoders}
\begin{document}
\sloppy
\author{
\begin{tabularx}{\textwidth}{X X}
Manpreet Singh Minhas & University of Waterloo, ON, Canada\\
John Zelek  & University of Waterloo, ON, Canada\\
\multicolumn{2}{l}{Email: \{msminhas,jzelek\}@uwaterloo.ca}
\end{tabularx}
}

\maketitle
\hyphenation{Humans}
\begin{abstract}
Anomaly detection refers to the task of finding unusual instances that stand out from the normal data. In several applications, these outliers or anomalous instances are of greater interest compared to the normal ones. Specifically in the case of industrial optical inspection and infrastructure asset management, finding these defects (anomalous regions) is of extreme importance. Traditionally and even today this process has been carried out manually. Humans rely on the saliency of the defects in comparison to the normal texture to detect the defects. However, manual inspection is slow, tedious, subjective and susceptible to human biases. Therefore, the automation of defect detection is desirable. But for defect detection lack of availability of a large number of anomalous instances and labelled data is a problem. In this paper, we present a convolutional auto-encoder architecture for anomaly detection that is trained only on the defect-free (normal) instances. For the test images, residual masks that are obtained by subtracting the original image from the auto-encoder output are thresholded to obtain the defect segmentation masks. The approach was tested on two data-sets and achieved an impressive average F1 score of 0.885. The network learnt to detect the actual shape of the defects even though no defected images were used during the training.
\end{abstract}

\section{Introduction}
An anomaly is anything that deviates from the norm. Anomaly detection refers to the task of finding the anomalous instances. Defect detection is a special case of anomaly detection and has applications in industrial settings. Manual inspection by humans is still the norm in most of the industries. The inspection process is completely dependent on the visual difference of the anomaly (defect) from the normal background or texture. The process is prone to errors and has several drawbacks, such as training time and cost, human bias and subjectivity, among others. Individual factors such as age, visual acuity, scanning strategy, experience, and training impact the errors caused during the manual inspection process  \cite{see_inspection}. As a result of these challenges faced in the manual inspection by humans, automation of defect detection has been a topic of research across different application areas such as steel surfaces \cite{app8112195}, rail tracks \cite{railtrack} and fabric \cite{fabricdefect}, to name a few. However, all these techniques face two common problems: lack of large labelled data and the limited number of anomalous samples. Semi-supervised techniques try to tackle this challenge. These techniques are based on the assumption that we have access to the labels for only one class type i.e. the normal class \cite{chandola2009anomaly}. They try to estimate the underlying distribution of the normal samples either implicitly or explicitly. This is followed by the measurement of deviation or divergence of the test samples from this distribution to determine an anomalous sample. To take an example of semi-supervised anomaly detection, Schlegl et al. \cite{anogan} used Generative Adversarial Networks (GANs) for anomaly detection in optical coherence tomography images of the retina. They trained a GAN on the normal data to learn the underlying distribution of the anatomical variability. But they did not train an encoder for mapping the input image to the latent space. Because of this, the method needed an optimization step for every test image to find a point in the latent space that corresponded to the most visually similar generated image which made it slow. In this research, we explore an auto-encoder based approach that also tries to estimate the distribution of the normal data and then uses residual maps to find the defects. It is described in the next section.

\section{Method}
\begin{figure*}[!ht]
\begin{center}
   \includegraphics[width=0.65\linewidth]{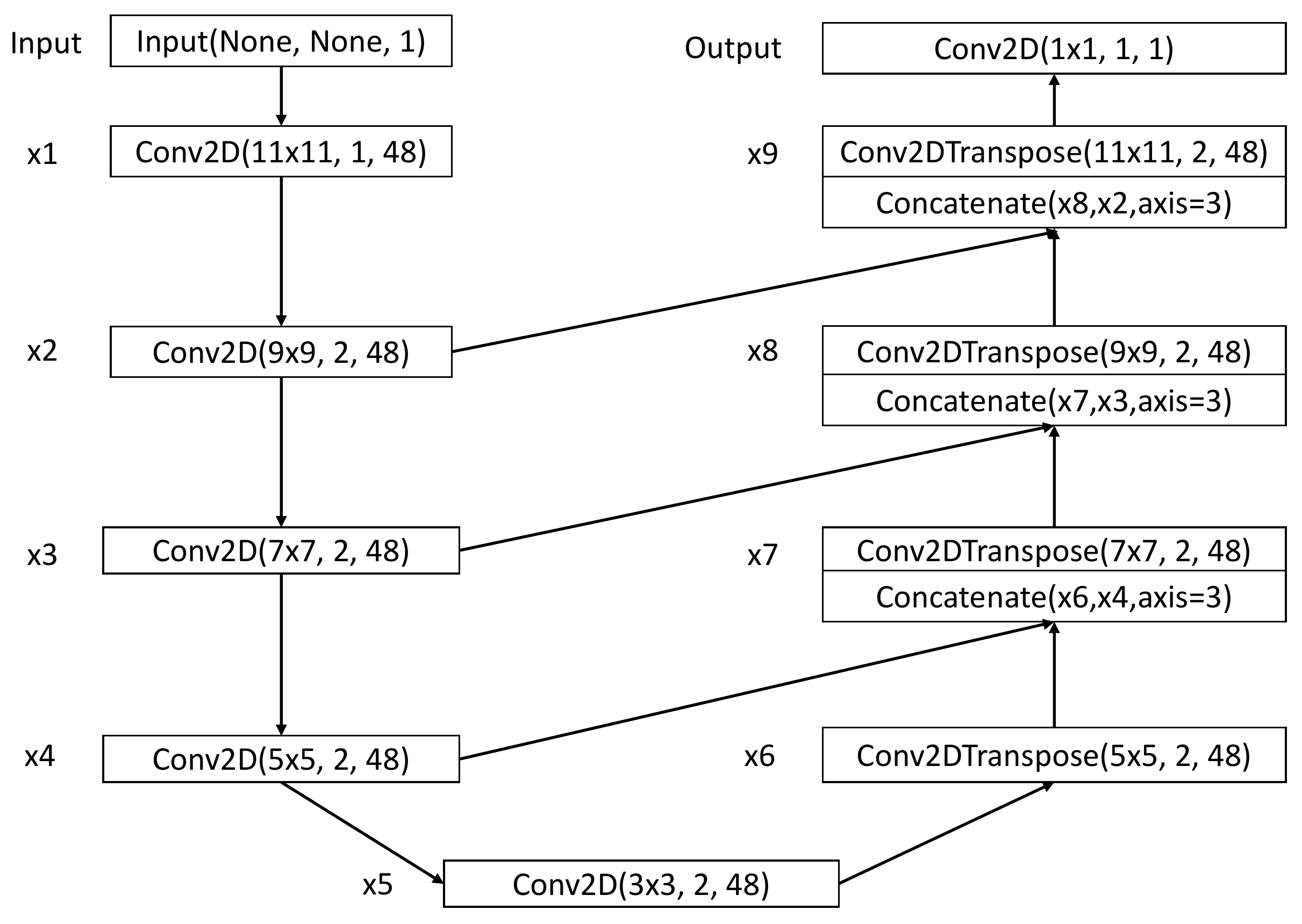}
\end{center}
   \caption{Proposed auto-encoder network architecture (similar to the UNet architecture). The encoder (layers x1 to x5) uses progressively decreasing filter sizes from $11 \times 11$ to $3 \times 3$. The decoder structure has kernel sizes that are in the reverse of the encoder order and uses Transposed Convolution Layers. The output from the encoder layers is concatenated with the previous layers before passing to layers x7 to x9. For every Conv2D(Transpose) layer the parameters shown are kernel size, stride and number of filters for that layer. After every layer, batch normalization is applied which is followed by the ReLU activation function. For a $H 
   \times W$ input the network outputs a $H 
   \times W$ reconstruction. The network is trained on only the defect-free or normal data samples.}
\label{fig:autoencoder_network}
\end{figure*}

The proposed network architecture is shown in Figure \ref{fig:autoencoder_network}. It is similar to the UNet \cite{unet} architecture. The encoder (layers x1 to x5) uses progressively decreasing filter sizes from $11 \times 11$ to $3 \times 3$. This decreasing filter size is chosen to allow for a larger field of view for the network without having to use large number of smaller size filters. Since deeper networks have a greater tendency to over-fit to the data and have poor generalization. The decoder structure has kernel sizes that are in the reverse of the encoder order and uses Transposed Convolution Layers. The output from the encoder layers is concatenated with the previous layers before passing to layers x7 to x9. For every Conv2D(Transpose) layer the parameters shown are kernel size, stride and number of filters for that layer. After every layer, batch normalization \cite{batchnormalization} is applied which is followed by the ReLU activation function \cite{Krizhevsky:2017:ICD:3098997.3065386}. For a $H 
   \times W$ input the network outputs a $H 
   \times W$ reconstruction. The network is trained on only the defect-free or normal data samples. Tensorflow 2.0 was used for conducting the experiments. The loss function used was the L2 norm or MSE (Mean Squared Error). The label in this case is the original input image and the prediction is the image reconstructed by the auto-encoder. Adam optimizer \cite{Adam} was used with default settings. The training was done for 50 epochs.

Our hypothesis is that the auto-encoder will learn representations that would only be able to encode and decode the normal samples properly and will not be able to reconstruct the anomalous regions. This shall cause large residuals for the defective regions in the residual map obtained by subtracting the reconstructed image from the input image as shown in Equation \ref{eq:residual}. The subtraction is done at per pixel-level. This is followed by a thresholding operation to obtain the final defect segmentation.

\begin{equation}
\label{eq:residual}
   R = X - AE(X)
\end{equation}
where $R$ is the residual, $X$ is the input and $AE(X)$ is the output (reconstructed image) of the auto-encoder. The data-sets used for conducting the experiments are described next.

\section{Data-sets}

\begin{enumerate}
    \item \label{ds:dagm} \textbf{DAGM}\cite{dagm} is a synthetic data-set for industrial optical inspection and contains ten classes of artificially generated textures with anomalies. For this study, the Class\;8 having the crack defect was randomly selected. It (hereafter referred to as DAGMC8) contains $150$ images with one defect per image and $1000$ defect-free images. 

    \item \textbf{RSDDs (Rail surface discrete defects)} \cite{raildefect} contains varying sized images of two different types of rails. We randomly selected the RSDDs Type-I category (referred to as RSDDsI) containing 67 images from express rails for the experiments. Segmentation masks were available which were used to extract $200 \times 160$ patches from the images and were classified into the anomaly and normal class to build the training and test data-set. 
\end{enumerate}

\begin{figure*}[!ht]
\begin{center}
   \includegraphics[width=0.7\linewidth]{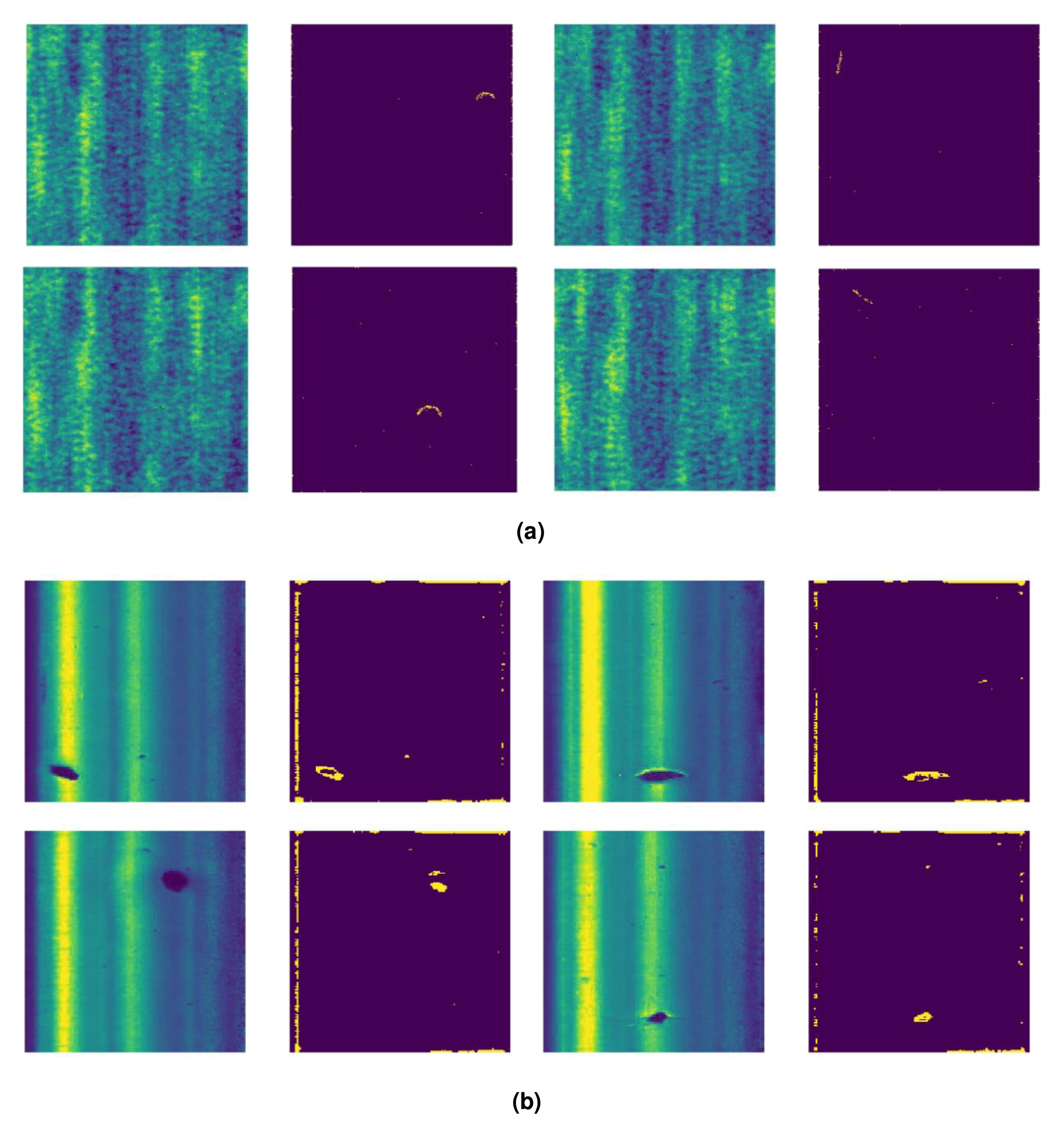}
\end{center}
   \caption{Few examples of defect detection output on the DAGMC8 (Fig. \ref{fig:results} (a)) and RSDDsI (Fig. \ref{fig:results} (b)) data-sets respectively.}
\label{fig:results}
\end{figure*}

\section{Results}
A few examples of segmentation results obtained after applying thresholding operation to the residual maps on the DAGMC8 and RSDDsI data-set are shown in \ref{fig:results} (a) and (b) respectively and the F1 score values are shown in Table \ref{tab:results}. As can be seen, on the synthetic DAGMC8 data-set the network could detect the cracks (anomalies) and there was little to no noise. This is in concurrence with the high F1 score value of 0.96. However, for the RSDDsI data-set, the results are a bit noisy. The same point is reflected by a lower F1 Score value of 0.81. Even though the thresholded residual maps managed to detect the defects in most of the images, the results contained more noise. One more observation was that the segmentation result was very sensitive to the choice of the threshold and minor changes led to large variations in the detection output. For the RSDDsI data-set, the illumination conditions were also varying in addition to the inherent noise in the data-set. For some images in the data-set, these areas were also not properly reconstructed by the auto-encoder, leading to false positives. Even though the geometric shapes and extent of the defects were different across images, the actual shapes of the anomalies were detected. This could be beneficial for applications where certain specific metrics need to be calculated for the defects. However, the lack of control over the types of defects that are detected by the auto-encoder reduces the targeting capability in comparison to supervised approaches.

\begin{table}[!ht]
    \centering
    \caption{F1 Score values on the DAGMC1 and RSDDsI data-set obtained by using the proposed method.}
    \begin{tabular}{ll}
    \toprule
        Data-set & F1 Score \\ \midrule
        DAGMC8   & 0.96    \\
        RSDDsI   & 0.81 \\ \bottomrule     
    \end{tabular}
    \label{tab:results}
\end{table}

\section{Conclusion and Future Work}
In this work, we explored and presented a semi-supervised anomaly detection technique using deep learning based AutoEncoders. The proposed network architecture is similar to UNet. It can be trained using only the normal samples. This is an important feature that is essential for practical applications where a limited number of anomalous samples and a large number of normal samples are available. The approach led to an impressive average F1 score of 0.885 on two data-sets. Qualitative results obtained on two data-sets show that the technique leads to the detection of anomalies which can vary in terms of shape, geometry, etc. However, the method is sensitive to the choice of the threshold. Even illumination changes were picked up by the method as anomalies which is undesirable. For future work, experiments on data-sets with more than one defect type per image could be conducted. Structural Similarity Index (SSIM) could be explored as a loss function. It compares two images based on luminance, contrast, and structure and as a result is a better measure of visual similarity in comparison to the mean squared error. Also, some kind of statistic such as the L1 norm calculated on the residual images could be used as an anomaly score. Rather than subtracting the reconstructed image, other comparison methods should be explored.  Exploring ways to make the network invariant to irrelevant factors such as illuminance needs to also be explored.

\section*{Acknowledgments}
We thank the Ontario Ministry of Transportation and NSERC (National Science and Research Council) for providing funds that supported this research.

\bibliographystyle{IEEEtran}  
\bibliography{references}

\begin{thebibliography}{10}
\providecommand{\url}[1]{#1}
\csname url@samestyle\endcsname
\providecommand{\newblock}{\relax}
\providecommand{\bibinfo}[2]{#2}
\providecommand{\BIBentrySTDinterwordspacing}{\spaceskip=0pt\relax}
\providecommand{\BIBentryALTinterwordstretchfactor}{4}
\providecommand{\BIBentryALTinterwordspacing}{\spaceskip=\fontdimen2\font plus
\BIBentryALTinterwordstretchfactor\fontdimen3\font minus
  \fontdimen4\font\relax}
\providecommand{\BIBforeignlanguage}[2]{{%
\expandafter\ifx\csname l@#1\endcsname\relax
\typeout{** WARNING: IEEEtran.bst: No hyphenation pattern has been}%
\typeout{** loaded for the language `#1'. Using the pattern for}%
\typeout{** the default language instead.}%
\else
\language=\csname l@#1\endcsname
\fi
#2}}
\providecommand{\BIBdecl}{\relax}
\BIBdecl

\bibitem{see_inspection}
\BIBentryALTinterwordspacing
J.~E. See, C.~G. Drury, A.~Speed, A.~Williams, and N.~Khalandi, ``The role of
  visual inspection in the 21st century,'' \emph{Proceedings of the Human
  Factors and Ergonomics Society Annual Meeting}, vol.~61, no.~1, pp. 262--266,
  2017. [Online]. Available: \url{https://doi.org/10.1177/1541931213601548}
\BIBentrySTDinterwordspacing

\bibitem{app8112195}
\BIBentryALTinterwordspacing
X.~Sun, J.~Gu, S.~Tang, and J.~Li, ``Research progress of visual inspection
  technology of steel products — a review,'' \emph{Applied Sciences}, vol.~8,
  no.~11, 2018. [Online]. Available:
  \url{http://www.mdpi.com/2076-3417/8/11/2195}
\BIBentrySTDinterwordspacing

\bibitem{railtrack}
H.~{Yu}, Q.~{Li}, Y.~{Tan}, J.~{Gan}, J.~{Wang}, Y.~{Geng}, and L.~{Jia}, ``A
  coarse-to-fine model for rail surface defect detection,'' \emph{IEEE
  Transactions on Instrumentation and Measurement}, vol.~68, no.~3, pp.
  656--666, March 2019.

\bibitem{fabricdefect}
A.~{Kumar}, ``Computer-vision-based fabric defect detection: A survey,''
  \emph{IEEE Transactions on Industrial Electronics}, vol.~55, no.~1, pp.
  348--363, Jan 2008.

\bibitem{chandola2009anomaly}
\BIBentryALTinterwordspacing
V.~Chandola, A.~Banerjee, and V.~Kumar, ``Anomaly detection: A survey,''
  \emph{ACM Computing Surveys (CSUR)}, vol.~41, no.~3, p.~15, 2009. [Online].
  Available:
  \url{http://scholar.google.de/scholar.bib?q=info:jAfBmk-9uAcJ:scholar.google.com/&output=citation&hl=de&ct=citation&cd=0}
\BIBentrySTDinterwordspacing

\bibitem{anogan}
T.~Schlegl, P.~Seeb{\"o}ck, S.~M. Waldstein, U.~Schmidt-Erfurth, and G.~Langs,
  ``Unsupervised anomaly detection with generative adversarial networks to
  guide marker discovery,'' in \emph{Information Processing in Medical
  Imaging}, M.~Niethammer, M.~Styner, S.~Aylward, H.~Zhu, I.~Oguz, P.-T. Yap,
  and D.~Shen, Eds.\hskip 1em plus 0.5em minus 0.4em\relax Cham: Springer
  International Publishing, 2017, pp. 146--157.

\bibitem{unet}
\BIBentryALTinterwordspacing
O.~Ronneberger, P.Fischer, and T.~Brox, ``U-net: Convolutional networks for
  biomedical image segmentation,'' in \emph{Medical Image Computing and
  Computer-Assisted Intervention (MICCAI)}, ser. LNCS, vol. 9351.\hskip 1em
  plus 0.5em minus 0.4em\relax Springer, 2015, pp. 234--241, (available on
  arXiv:1505.04597 [cs.CV]). [Online]. Available:
  \url{http://lmb.informatik.uni-freiburg.de/Publications/2015/RFB15a}
\BIBentrySTDinterwordspacing

\bibitem{batchnormalization}
\BIBentryALTinterwordspacing
S.~Ioffe and C.~Szegedy, ``Batch normalization: Accelerating deep network
  training by reducing internal covariate shift,'' in \emph{Proceedings of the
  32Nd International Conference on International Conference on Machine Learning
  - Volume 37}, ser. ICML'15.\hskip 1em plus 0.5em minus 0.4em\relax JMLR.org,
  2015, pp. 448--456. [Online]. Available:
  \url{http://dl.acm.org/citation.cfm?id=3045118.3045167}
\BIBentrySTDinterwordspacing

\bibitem{Krizhevsky:2017:ICD:3098997.3065386}
\BIBentryALTinterwordspacing
A.~Krizhevsky, I.~Sutskever, and G.~E. Hinton, ``Imagenet classification with
  deep convolutional neural networks,'' \emph{Commun. ACM}, vol.~60, no.~6, pp.
  84--90, May 2017. [Online]. Available:
  \url{http://doi.acm.org/10.1145/3065386}
\BIBentrySTDinterwordspacing

\bibitem{Adam}
\BIBentryALTinterwordspacing
D.~P. Kingma and J.~Ba, ``Adam: A method for stochastic optimization,'' 2014,
  cite arxiv:1412.6980Comment: Published as a conference paper at the 3rd
  International Conference for Learning Representations, San Diego, 2015.
  [Online]. Available: \url{http://arxiv.org/abs/1412.6980}
\BIBentrySTDinterwordspacing

\bibitem{dagm}
T.~H. Matthias~Wieler, ``Weakly supervised learning for industrial optical
  inspection,'' \url{https://hci.iwr.uni-heidelberg.de/node/3616}.

\bibitem{raildefect}
J.~{Gan}, Q.~{Li}, J.~{Wang}, and H.~{Yu}, ``A hierarchical extractor-based
  visual rail surface inspection system,'' \emph{IEEE Sensors Journal},
  vol.~17, no.~23, pp. 7935--7944, Dec 2017.

\end{thebibliography}
\end{document}